# Twin reflections from a moving space-time boundary

Yukun Yang[1,†], Hao Hu[1,†,*], Youxiu Yu[1], Linyang Zou[2], Liangliang Liu[1], Jiang Xiong[3], Baile Zhang[4,5], Francisco J. Garcia-Vidal[6], Zhuo Li[1,*] and Yu Luo[1,*]

[1]*National Key Laboratory of Microwave Photonics, College of Electronic and Information Engineering, Nanjing University of Aeronautics and Astronautics, Nanjing 211106, China*

[2]*School of Electrical and Electronic Engineering, Nanyang Technological University, Singapore 639798, Singapore*

[3]*Computational Electromagnetics Laboratory, School of Physics, University of Electronic Science and Technology of China, Chengdu 610054, China*

[4]*School of Physical and Mathematical Sciences, Nanyang Technological University, Singapore 637371, Singapore*

[5]*Centre for Disruptive Photonic Technologies, Nanyang Technological University, Singapore 639798, Singapore*

[6]*Departamento de Física Teórica de la Materia Condensada and Condensed Matter Physics Center (IFIMAC), Universidad Autónoma de Madrid, Madrid E-28049, Spain*

[†]These authors contribute equally.

[*]Corresponding authors: Hao Hu; Zhuo Li; Yu Luo

**Email:** hao.hu@nuaa.edu.cn (Hao Hu); lizhuo@nuaa.edu.cn (Zhuo Li); yu.luo@nuaa.edu.cn (Yu Luo)

**An interluminal interface, a space-time boundary propagating at a velocity between the group velocities of the surrounding media, enables extraordinary wave phenomena such as nonreciprocal amplification and analogues of Hawking radiation. A particularly intriguing prediction is that such an interface splits an incident wave into three outgoing waves: one transmitted and two reflected. Yet, despite decades of theoretical study, this triple-wave scattering has remained experimentally unobserved, owing to stringent requirements on interface velocity and modulation speed. Here, we introduce a programmable spatiotemporal microstrip transmission-line platform that realizes step-modulated interluminal interfaces with controlled velocity. Using this system, we report the direct observation of bi-reflection from an interluminal interface, confirming the emergence of two distinct reflected waves alongside a transmitted one. Measured frequencies and scattering coefficients show excellent agreement with longstanding theoretical predictions. Furthermore, we reveal that the two reflections possess fundamentally different causal symmetries: one is spatially inverted, while the other is time-reversed. These findings resolve a half-century-old puzzle in moving-boundary electrodynamics and establish a versatile experimental platform for studying wave interaction with dynamic interfaces. Our work opens pathways to velocity-independent scattering devices, broadband frequency conversion, and advanced spatiotemporal wave engineering.**

## 1. Introduction

The control of wave propagation through spatial inhomogeneity (governed by the Fresnel equations and Snell's law[1]) has long laid the foundation for modern photonics and metamaterials, enabling antireflection coatings[2], coherent wave manipulation[3], and invisibility cloaking[4]. In parallel, time has emerged as a powerful additional degree of freedom for wave control[5]. This temporal control can be achieved through a temporal interface, created by a sudden change in material properties in time. In contrast to a spatial interface which conserves energy but alters momentum, the temporal interface breaks temporal translational symmetry, permitting energy exchange between the wave and the medium and giving rise to distinctive scattering phenomena such as temporal reflection and frequency translation[6,7].

Recent advances in ultrafast optics and nanofabrication have spurred renewed interest in time-varying media[8–16], leading to experimental demonstrations in optical fibres[17], epsilon-near-zero films[18,19], and other platforms[6,20–28]. Among the rich phenomena explored, one of the most intriguing is the interaction of waves with an interface that moves uniformly through space and time. Such a "spatiotemporal interface" can be classified into three regimes depending on its velocity $v_b$ relative to the wave group velocities in the two adjacent media[29]: subluminal regime ($|v_b|$ below both group velocities), superluminal regime ($|v_b|$ above both), and interluminal regime ($|v_b|$ between the two). Crucially, the interluminal regime is predicted to host a wealth of counterintuitive effects[30–36], including broadband nonreciprocal amplification[33] and analog Hawking radiation[34,35].

Perhaps the most striking prediction for a step-modulated interluminal interface is the splitting of an incident wave into three outgoing waves: one transmitted and two reflected. The coexistence of two reflected waves is physically permitted since the interluminal interface velocity lies between the reflection velocities in the two media: it is slower than the reflected wave in the incident medium with lower refractive index, while exceeding that in the transmitted medium with higher refractive index. Later, we will denote such an anomalous phenomenon of twin reflections as bi-reflection. This bi-reflection was first identified as a boundary-condition paradox[37] in 1967, and later resolved theoretically in 1975 by introducing a virtual gradient-index transition at the interface[38]. Yet despite its fundamental importance and conceptual elegance, experimental verification has remained elusive for over half a century, demanding both precise control of the interface velocity within a narrow interluminal window and an ultrafast modulation response far shorter than the wave period. This long-standing gap between theory and experiment has left open basic questions regarding the actual scattering coefficients, frequency relations, and underlying symmetries of the reflected waves.

Here, we close this gap by realizing a fully programmable spatiotemporal microstrip transmission-line (STMTL) platform that enables the first experimental observation of bi-reflection at an interluminal interface. By dynamically toggling an array of subwavelength switch-loaded capacitors via a field-programmable gate array (FPGA), we achieve precise, tunable control of the interface velocity across subluminal, superluminal, and interluminal regimes. Using this system, we directly measure the three outgoing waves generated from a step-like interluminal interface. Their frequencies and scattering magnitudes show excellent agreement with the theoretical predictions formulated decades ago. Furthermore, we reveal that the two reflected waves possess distinct causal symmetries: one is spatially inverted while preserving the temporal sequence, whereas the other is time-reversed while maintaining the spatial sequence. Our work not only resolves a classic puzzle in moving-boundary electrodynamics but also establishes a versatile experimental testbed for studying wave interaction with arbitrarily programmed spatiotemporal interfaces, providing new insights into wave manipulation in spatiotemporal metamaterials.

## 2. Results and discussion

To begin with, we revisit the wave-matter interaction process in a system of interluminal interfaces. Without loss of generality, we consider a moving interface separating two media with refractive indices of $n_1$ and $n_2$, denoted as medium 1 and medium 2, respectively [see Fig. 1(a)]. This interface moves along the negative *x*-axis at velocity $v_b$. For concise, we use the normalized interface velocity $\beta = v_b/c$ below, where *c* is the light speed in vacuum. The interface is classified as an interluminal one when $v_2/c < |\beta| < v_1/c$, where $v_1 = c/n_1$ ($v_2 = c/n_2$) represents the group velocity of waves in medium 1 (medium 2). When a forward-propagating incident wave ($\psi_{i,1}$) interacts with the interluminal interface, three outgoing waves are theoretically permitted, namely, one transmitted wave ($\psi_{t,2}$) and two reflected waves ($\psi_{r,1}$ and $\psi_{r,2}$). Specifically, the transmitted wave $\psi_{t,2}$ propagates forward in medium 2 while the reflected waves $\psi_{r,1}$ and $\psi_{r,2}$ propagates backward in medium 1 and medium 2, respectively. This scattering behavior is also captured by the finite-difference time-domain (FDTD) simulations shown in Fig. 1(c). Enforcing the phase-matching conditions at the moving interface, the frequencies of the three outgoing waves $f_{r,1}$, $f_{r,2}$ and $f_{t,2}$ are derived as (see detailed derivations in Supplementary Information S4)

$$f_{r,1} = \frac{1 - n_1\beta}{1 + n_1\beta} f_i, \;\; f_{r,2} = \frac{1 - n_1\beta}{1 + n_2\beta} f_i, \;\; f_{t,2} = \frac{1 - n_1\beta}{1 - n_2\beta} f_i, \tag{1}$$

where $f_i$ is the frequency of the incident wave. Following the approach in Ref. [38], the reflection coefficients $R_{r,1}$, $R_{r,2}$ and the transmission coefficient $T_{t,2}$ are given by,

$$R_{r,1} = -1, \;\; R_{r,2} = -\frac{n_1}{n_2}, \;\; T_{t,2} = \frac{n_1}{n_2}. \tag{2}$$

Surprisingly, these scattering coefficients are independent of the interface velocity, whereas in the subluminal and the superluminal regimes, the reflection and transmission coefficients exhibit explicit velocity dependence (see Supplementary Information S2&S3).

For practical implementation, we adopt a platform of STMTL constructed by a microstrip transmission-line (MTL) with a switch-controlled capacitor $C_{\text{load}}$ loaded in parallel. The subwavelength period of unit cell is $\Delta x = 0.208$ m [see Fig. 1(b)]. The MTL has parameters of relative dielectric constant $\varepsilon_r = 10.8$, substrate thickness $h = 1.28$ mm, and metal patch width $w = 1.1$ mm [see the inset of Fig. 1(b)]. When the capacitors are unloaded, the effective refractive index of the MTL (medium 1) is $n_1 = 2.67$ (see more details in Supplementary Information S6). Such a refractive index is effectively changed by loading capacitors onto the MTL. Here, the adopted capacitance is $C_{\text{load}} = 82$ pF, leading to the effective refractive index $n_2 = 4.74$ of medium 2. To realize the moving interface, we utilize FPGA to trigger on the switches with an equal time step $\Delta t$ from port 2 to port 1 in sequence. Such a method enables us to reach the interluminal regime. For example, when $\Delta t = 2.82$ ns, $\beta = -0.246$ satisfies the interluminal conditions in our systems (i.e., $v_2/c = 0.211 < |\beta| < v_1/c = 0.375$). By constructing the circuit model with the above parameters, we simulate the voltage variations at port 1 and port 2 [see Fig. 1(d&e)] in an event of spatiotemporal scattering with an interluminal interface. In this simulation, we input a Gaussian pulse $\psi_{i,1}$ from port 1, and set the normalized velocity of interluminal interface as $\beta = -0.246$. From the simulated result, two reflected signals $\psi_{r,1}$ (red) and $\psi_{r,2}$ (blue) occur at port 1 [see Fig.1 (d)], and one transmitted signal $\psi_{t,2}$ (green) occurs at port 2 [see Fig.1 (e)]. The simulated voltage waves at input and output ports in STMTL are in excellent agreement with those using FDTD simulations [Fig. 1(c)], demonstrating the feasibility of our proposed platform to realize the bi-reflection effect.

To observe the phenomenon of three scattered waves in reality, we fabricate a sample of the STMTL system [see Fig. 1(f)]. The meandered MTL is printed on a 1.28 mm thick Rogers RO3210 substrate (see more specific parameters and details in the caption of Fig. 1 and the methods section). The switch modules are integrated onto the substrate, and the control terminals of these switches [i.e., the port of transistor-transistor logic input (TTL-IN)] are connected to the FPGA (see more details in Supplementary Information S1). In the experiment, we use an arbitrary waveform generator (AWG) to generate a Gaussian pulse $\psi_{i,1}$, and input $\psi_{i,1}$ into the MTL from port 1. Meanwhile, we utilize FPGA to trigger switches with $\Delta t = 2.58$ ns, to form an interluminal interface with $\beta = -0.269$. We repeat the above operations to realize identical scattering events periodically. Moreover, we use an oscilloscope to record the voltage variations at port 1 and port 2 during these scattering events. Fig. 2 (a&b) shows a group of test waveforms of spatiotemporal scatterings in the interluminal regime. To obtain convincing results, we use the oscilloscope to sample 256 groups of waveforms and average them. The experimental measurement results clearly reflect three scattered waves: the two reflected waves $\psi_{r,1}$ (as highlighted in red) is detected at 142 ns and $\psi_{r,2}$ (as highlighted in blue) is detected at 160 ns [see Fig. 2(a)], one transmitted wave $\psi_{t,2}$ (as highlighted in green) is detected at 79 ns [see Fig. 2(b)].

Our measured frequencies and magnitudes of three scattered waves are almost overlapped with theoretical and numerical ones in the interluminal regime. In the case of interluminal regime, one reflected wave $\psi_{r,1}$ occurs on the left side of the interface, while the other one $\psi_{r,2}$ occurs on the right side of the interface [see the middle inset in Fig. 2(h)]. To qualify the measured frequencies and the scattering coefficients of three scattered waves, we perform the Fourier transform of the waveforms [see Fig. 2(c~f)]. The measured frequencies $f_{\text{meas.}}$ of scattered waves are obtained by extracting central frequencies of corresponding waveforms. Compared to the incident frequency $f_{i,1_\text{meas.}} = 18.5$ MHz, the frequencies of two reflected waves exhibit a noticeable enhancement, i.e., $f_{r,1_\text{meas.}} = 100.8$ MHz ($f_{r,1_\text{theory}} = 113.06$ MHz) and $f_{r,2_\text{meas.}} = -104$ MHz ($f_{r,2_\text{theory}} = -115.2$ MHz), where $f_{\text{theory}}$ are the theoretically predicted frequencies from Eq. (1). On the other hand, the frequency of the transmitted wave decreases as $f_{t,2_\text{meas.}} = 13.75$ MHz ($f_{t,2_\text{theory}} = 13.97$ MHz). The scattering coefficients of three scattered waves are measured by normalizing the corresponding magnitudes of waveforms at the central frequencies to that of the incidence (see more details in Supplementary Information S5). To be specific, the measured reflection coefficients of $\psi_{r,1}$ and $\psi_{r,2}$ are $R_{1_\text{meas.}} = -0.7611$ ($R_{1_\text{theory}} = -1$) and $R_{2_\text{meas.}} = -0.6652$ ($R_{2_\text{theory}} = -0.563$), respectively; the measured transmission coefficient of $\psi_{t,2}$ is $T_{\text{meas.}} = 0.7518$ ($T_{\text{theory}} = 0.563$). Here, $R(T)_{\text{theory}}$ is the theoretically predicted reflection (transmission) coefficients from Eq. (2). Following the above approach, we measured three groups of scattering coefficients and frequencies of the corresponding scattered waves at different velocities of the interluminal interface [see Fig. 2(g&h)]. Our experimentally measured frequencies of these scattered waves are consistent with theoretical predictions, showing strong dependence on the interface velocity. On the other hand, while the theoretical results reflect that the magnitudes of the scattered waves are irrelevant to the interface velocity, the experimentally measured ones are oscillated slightly over the theoretical values especially for two reflections. These deviations are probably because the extremely high frequencies of reflected waves invalidate the effective medium theory as the interface velocity $\beta$ approaches the phase velocity of waves in the surrounding media, namely, $v_1/c$ or $v_2/c$. Moreover, recent advances suggest that such deviations may also be related to Hawking radiation[35]. To be specific, the interluminal interface forms an optical event horizon. This horizon amplifies quantum vacuum fluctuations. Such an effect could lead to stimulated Hawking radiation, which is not considered in theory or simulation and worths future investigation.

In addition to interluminal regime, our measured frequencies and magnitudes of scattered waves are in excellent agreement with the theoretical and numerical results in subluminal and superluminal regimes. Different from the case in the interluminal regime, only one reflection and one transmission are permitted in subluminal and superluminal regimes. In the subluminal regime with $|\beta| < v_2/c = 0.211$, the reflection and transmission occur on different sides of the subluminal interface [see the rightmost inset in Fig. 2(h)]. In superluminal with $|\beta| > v_1/c = 0.375$, the reflection and transmission occur on the same side of the superluminal interface [see the leftmost inset in Fig. 2(h)]. Despite aforementioned difference, our theoretical, numerical and experimental results [see Fig. 2(g&h)] consistently show that the frequency and magnitude of reflection and transmission in subluminal and superluminal regimes highly depend on the interface velocity.

Finally, we find that two reflected waves from the interluminal interface exhibit different causalities [see Fig. 3(a)]. To demonstrate this difference, we impose a pair of asymmetric Gaussian pulses $\psi_{i,1}$ into the STMTL. The first pulse has a higher magnitude (as marked by the pink dot), while the second one has a lower magnitude (as marked by the black dot). The first reflection $\psi_{r,1}$ received by port 1 at 80 ns [see red envelope in Fig. 3(b)] preserves the same sequence as the incidence in time as the pulse with the higher magnitude arrives first, followed by the pulse with a lower magnitude. This reveals that the $\psi_{r,1}$ is spatial-inverted. The second reflection $\psi_{r,2}$ received by port 1 at 93 ns [see blue envelope in Fig. 3(b)] exhibits an opposite sequence as the incidence in time, where the pulse with the lower magnitude reaches port 1 first, followed by the pulse with a higher magnitude. This phenomenon reflects the time-reversed nature of $\psi_{r,2}$. On the other hand, the spatial and temporal sequences of the transmission $\psi_{t,2}$ received by port 2 at 100 ns are identical to the incidence [see green envelope in Fig. 3(c)]. In this way, Fig. 3 highlights the interluminal interface as a symmetry-selective scatterer, producing reflected waves with fundamentally different causalities.

**3. Conclusion**

In summary, we have developed a fully programmable STMTL platform that enables precise control over the velocity of a step-modulated interface across subluminal, superluminal, and interluminal regimes. Using this system, we report the first experimental observation of bi-reflection at an interluminal interface, thereby resolving a half-century-old theoretical puzzle in moving-boundary electrodynamics. Our measurements confirm not only the predicted triple-wave output (one transmitted and two reflected waves) but also validate the remarkable velocity-invariance of their scattering coefficients, in excellent agreement with the model proposed in 1975.

Beyond amplitude and frequency verification, we uncover a fundamental causal distinction between the two reflected waves: one exhibits spatial inversion while preserving temporal sequence, whereas the other is time-reversed while retaining spatial order. This symmetry-selective scattering reveals deeper wave-manipulation mechanisms inherent to spatiotemporal interfaces. The velocity-independent reflection coefficient offers a promising basis for counter-reconnaissance systems resilient to Doppler-based detection, while the interface's ability to amplify arbitrary frequencies as it approaches the phase velocity of the host medium surpasses conventional nonlinear conversion limits. We emphasize that the findings reported here are general in nature and broadly extendable. First, they could be translated into the microwave, terahertz and even optical bands by implementing advanced photonic platforms such as high-speed photodiode-based metasurface[20], time-synthetic lattice with dual coupled fibre loops[39], and a broad class of nonlinear media[19]. Second, our results are expected to stimulate rapid progress in interluminal wave manipulation across neighboring disciplines, such as acoustics[40], elastodynamics[23], and quantum mechanics[41].

Beyond these immediate extensions, our platform establishes a versatile and reconfigurable testbed for studying wave dynamics under programmable spatiotemporal modulations. This capability paves the way for advanced control paradigms such as spatiotemporal coding[42], topological pumping[41], and tabletop realizations of gravitational analog metrics[43]. Such a technical advance not only closes a long-standing experimental gap but also unlocks a practical pathway to explore richer wave-interface physics, from accelerated boundaries that emulate curved spacetime to non-Hermitian and quantum-informed spatiotemporal metamaterials.

**Reference**

1. Papas, C. H. *Theory of Electromagnetic Wave Propagation*. (Courier Corporation, 2014).
2. Pozar, D. M. *Microwave Engineering: Theory and Techniques*. (John Wiley & Sons, 2021).
3. Baranov, D. G., Krasnok, A., Shegai, T., Alù, A. & Chong, Y. Coherent perfect absorbers: linear control of light with light. *Nat. Rev. Mater.* **2**, (2017).
4. Pendry, J. B., Schurig, D. & Smith, D. R. Controlling Electromagnetic Fields. *Science* **312**, 1780–1782 (2006).

5. Moshinsky, M. Diffraction in Time. *Phys. Rev.* **88**, 625–631 (1952).
6. Moussa, H. *et al.* Observation of temporal reflection and broadband frequency translation at photonic time interfaces. *Nat. Phys.* **19**, 863–868 (2023).
7. Pendry, J. B. Time Reversal and Negative Refraction. *Science* **322**, 71–73 (2008).
8. Yu, L. *et al.* Dirac mass induced by optical gain and loss. *Nature* **632**, 63–68 (2024).
9. Schiller, O., Plotnik, Y., Segal, O., Lyubarov, M. & Segev, M. Time-Domain Bound States in the Continuum. *Phys. Rev. Lett.* **133**, 263802 (2024).
10. Lyubarov, M. *et al.* Amplified emission and lasing in photonic time crystals. *Science* **377**, 425–428 (2022).
11. Yang, Y. *et al.* Topologically protected edge states in time photonic crystals with chiral symmetry. *ACS Photonics* **12**, 2389–2396 (2025).
12. Dikopoltsev, A. *et al.* Light emission by free electrons in photonic time-crystals. *Proc. Natl. Acad. Sci. U.S.A.* **119**, e2119705119 (2022).
13. Pan, Y., Cohen, M.-I. & Segev, M. Superluminal k -Gap Solitons in Nonlinear Photonic Time Crystals. *Phys. Rev. Lett.* **130**, 233801 (2023).
14. Bae, J., Lee, K., Min, B. & Kim, K. W. Quantum electrodynamics of photonic time crystals. *Nat. Commun.* **17**, 0–12 (2025).
15. Yu, Y. *et al.* Antireflection Spatiotemporal Metamaterials. *Laser Photonics Rev.* **17**, 2300130 (2023).
16. Yu, Y. *et al.* Generalized coherent wave control at dynamic interfaces. *Laser Photonics Rev.* **19**, 2400399 (2025).
17. He, R. *et al.* Integration of gigahertz-bandwidth semiconductor devices inside microstructured optical fibres. *Nat. Photonics* **6**, 174–179 (2012).
18. Tirole, R. *et al.* Double-slit time diffraction at optical frequencies. *Nat. Phys.* **19**, 999–1002 (2023).
19. Alam, M. Z., De Leon, I. & Boyd, R. W. Large optical nonlinearity of indium tin oxide in its epsilon-near-zero region. *Science* **352**, 795–797 (2016).
20. Jones, T. R., Kildishev, A. V., Segev, M. & Peroulis, D. Time-reflection of microwaves by a fast optically-controlled time-boundary. *Nat. Commun.* **15**, 6786 (2024).
21. Dong, Z. *et al.* Quantum time reflection and refraction of ultracold atoms. *Nat. Photonics* **18**, 68–73 (2024).
22. Bacot, V., Labousse, M., Eddi, A., Fink, M. & Fort, E. Time reversal and holography with spacetime transformations. *Nat. Phys.* **12**, 972–977 (2016).
23. Delory, A. *et al.* Elastic Wave Packets Crossing a Space-Time Interface. *Phys. Rev. Lett.* **133**, 267201 (2024).
24. Lee, K. *et al.* Linear frequency conversion via sudden merging of meta-atoms in time-variant metasurfaces. *Nat. Photonics* **12**, 765–773 (2018).
25. Xiong, J. *et al.* Observation of wave amplification and temporal topological state in a non-synthetic photonic time crystal. *Nat. Commun.* **16**, 11182 (2025).
26. Wang, X. *et al.* Metasurface-based realization of photonic time crystals. *Sci. Adv.* **9**, eadg7541 (2023).
27. Hou, H. *et al.* Experimental realization of a full-band wave antireflection based on temporal taper metamaterials. *Commun. Phys.* 0–25 (2026).
28. Ren, Y. *et al.* Observation of momentum-gap topology of light at temporal interfaces in a time-synthetic lattice. *Nat. Commun.* **16**, 707 (2025).
29. Deck-Léger, Z.-L., Chamanara, N., Skorobogatiy, M., Silveirinha, M. G. & Caloz, C. Uniform-velocity spacetime crystals. *Adv. Photonics* **1**, 1–27 (2019).

30. Dinc, T. *et al.* Synchronized conductivity modulation to realize broadband lossless magnetic-free non-reciprocity. *Nat. Commun.* **8**, 795 (2017).
31. Yang, Q. *et al.* Spatiotemporal superfocusing. *Laser Photonics Rev.* **19**, 2500126 (2025).
32. Wu, H. *et al.* Conformal Spatiotemporal Modulation Enabled Geometric Frequency Combs. *ACS Photonics* **11**, acsphotonics.4c00029 (2024).
33. Galiffi, E., Huidobro, P. A. & Pendry, J. B. Broadband Nonreciprocal Amplification in Luminal Metamaterials. *Phys. Rev. Lett.* **123**, 206101 (2019).
34. Pendry, J. B., Galiffi, E. & Huidobro, P. A. Photon conservation in trans-luminal metamaterials. *Optica* **9**, 724 (2022).
35. Horsley, S. A. R. & Pendry, J. B. Quantum electrodynamics of time-varying gratings. *Proc. Natl. Acad. Sci. U.S.A.* **120**, e2302652120 (2023).
36. Wu, H. *et al.* Electromagnetic wave trapping at the discretized interluminal interface. *Laser Photonics Rev.* e00975 (2025).
37. Ostrovskii, L. A. & Solomin, B. A. Correct formulation of the problem of wave interaction with a moving parameter jump. *Radiophys. Quant. Electron.* **10**, 666–668 (1967).
38. Ostrovskii, L. A. Some ‘moving boundaries paradoxes’ in electrodynamics. *Sov. Phys. Usp.* **18**, 452–458 (1975).
39. Ren, Y. *et al.* Observation of momentum-gap topology of light at temporal interfaces in a time-synthetic lattice. *Nat. Commun.* **16**, 707 (2025).
40. Chen, Z. *et al.* Efficient nonreciprocal mode transitions in spatiotemporally modulated acoustic metamaterials. *Sci. Adv.* **7**, eabj1198 (2021).
41. Feis, J., Weidemann, S., Sheppard, T., Price, H. M. & Szameit, A. Space-time-topological events in photonic quantum walks. *Nat. Photonics* **19**, 518–525 (2025).
42. Zhang, L. *et al.* Space-time-coding digital metasurfaces. *Nat. Commun.* **9**, 4334 (2018).
43. Pendry, J. B. & Horsley, S. A. R. QED in space–time varying materials. *APL Quantum* **1**, 20901 (2024).

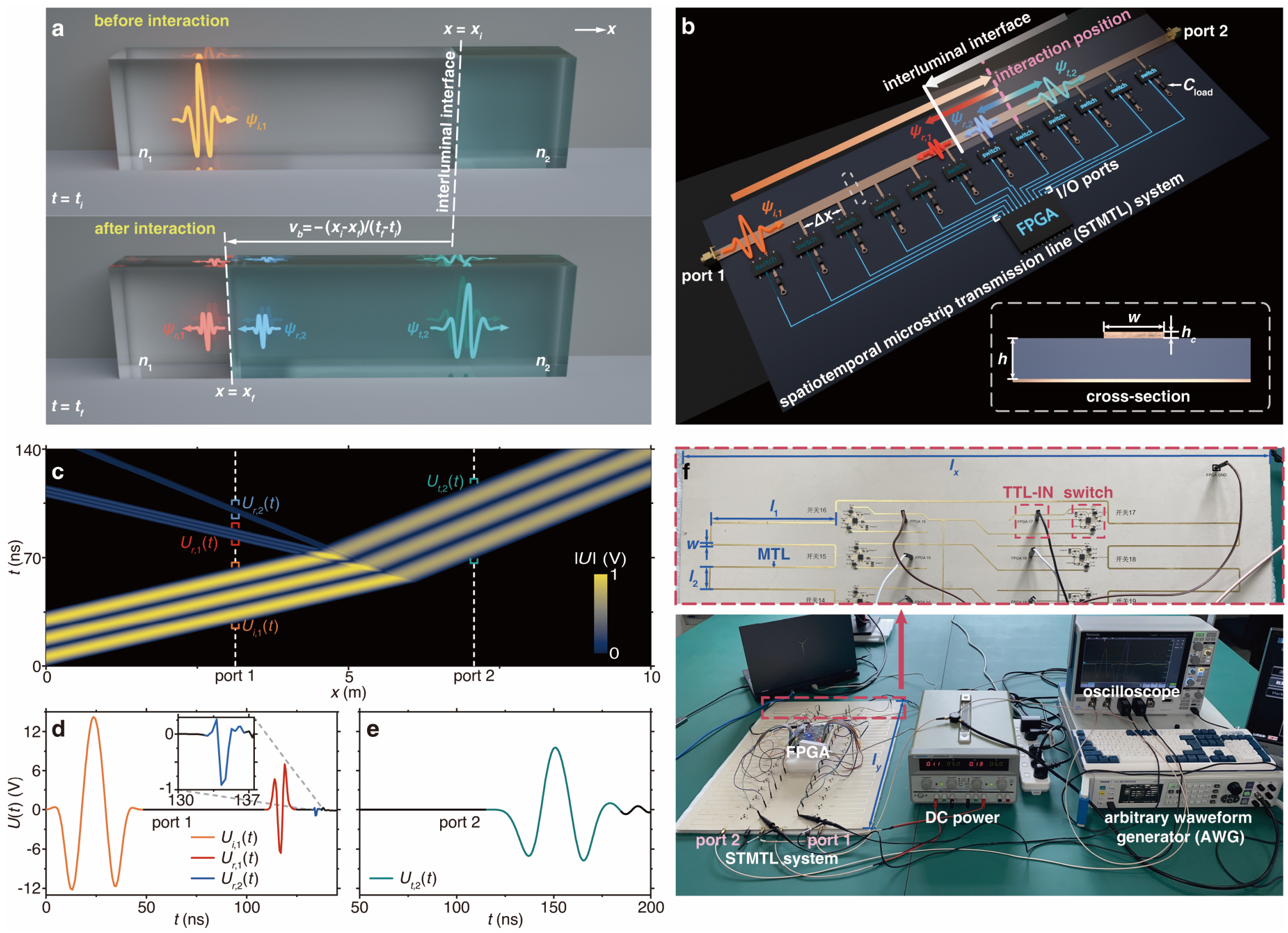

**Fig. 1 | Concept of bi-reflection effect at interluminal interface and experimental setup of the STMTL.** (a) The schematic of the bi-reflection effect at interluminal interface. The incident wave $\psi_{i,1}$ (orange) propagates along the $x$-axis. The interluminal interface (white dotted line) moves along the negative $x$-axis, with the velocity $v_b = -(x_i - x_f)/(t_f - t_i)$, where $x_i(x_f)$ is the interface position at the moment $t_i(t_f)$. The interaction of the incident wave and interluminal interface forms three scattered waves, including two scattered waves $\psi_{r,1}$, $\psi_{r,2}$ and one transmitted wave $\psi_{t,2}$. The two reflected waves propagate in different media ($\psi_{r,1}$ in medium 1 with refractive index $n_1$ and $\psi_{r,2}$ in medium with refractive index $n_2$), while the transmitted wave $\psi_{t,2}$ propagates in medium 2. (b) Schematic of the STMTL system. The incident signal $\psi_{r,1}$ is input from port 1, propagating along the main part of the MTL (the propagating direction is highlighted with an orange arrow). The illustration shows the parameters of the MTL (with substrate thickness $h = 1.28$ mm, metal patch width $w = 1.1$ mm, patch thick $h_c = 0.035$ mm, and period of unit cell $\Delta x = 0.208$ m). FPGA triggers the switches in sequence from port 2 to port 1, forming an interluminal interface [the moving direction is highlighted with the white arrow, and the bright (dark) switches are under the on (off) state]. $\psi_{r,1}$ travels to the position indicated by the pink dashed line and starts the interaction with the interface. After a short duration, the interface moves to the position marked with white line, accompanied by the generation of three scattered waves. (c) FDTD simulation of voltage-wave distribution in the system of interluminal interface. (d,e) Simulated voltage waves at input and output ports in STMTL. In (d), the incident voltage wave $U_{i,1}(t)$ is marked in orange, the reflected voltage wave $U_{r,1}(t)$ [$U_{r,2}(t)$] is marked in blue (red). The transmitted voltage wave is marked in green. (f) The photograph of STMTL and the measurement system. The AWG is connected with port 1 of the STMTL sample to input a Gaussian pulse into the system. The MTL sample has board length $l_x = 610$ mm and broad width $l_y = 457$ mm. The parameters of the MTL are $l_1 = 88$ mm, $l_2 = 16$ mm, and $w = 1.1$ mm (see red dashed box). The I/O ports of the FPGA are connected to the control pin (TTL-IN) of the fast switches powered by a DC power. We connect port 1 and port 2 of the STMTL with an oscilloscope to capture the waveforms of the scattered waves.

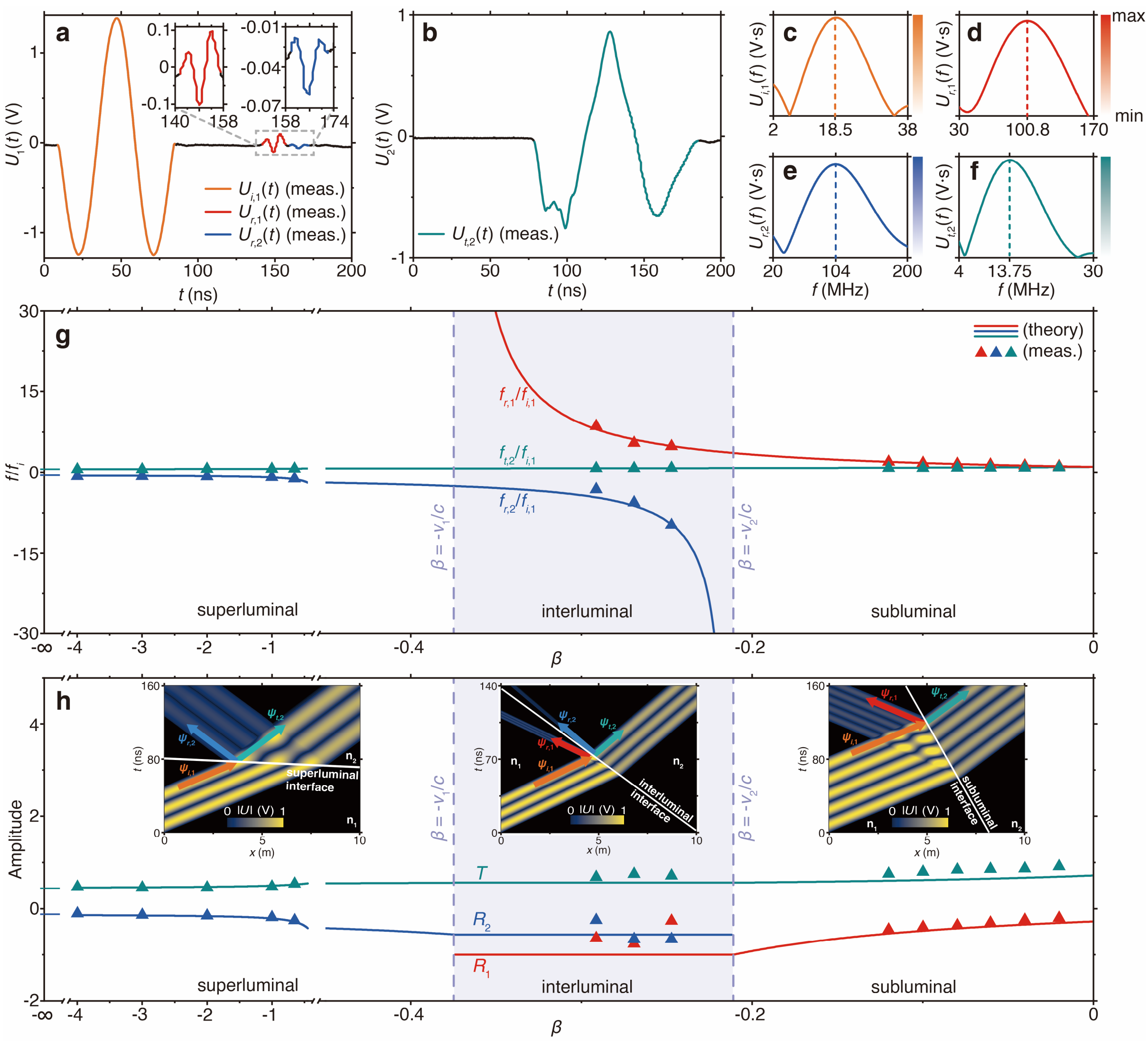

**Fig. 2 | Spectral analysis of spatiotemporal scattering.** (a,b) Measured waveforms of spatiotemporal scattering in the interluminal regime (corresponding normalized interface velocity $\beta = -0.269$). (a) Measured waveform at port 1 [incident voltage wave $U_{i,1}(t)$, reflected voltage waves $U_{r,1}(t)$ and $U_{r,2}(t)$]. (b) Measured waveform at port 2 [transmitted voltage wave $U_{t,2}(t)$]. (c~f) The Fourier transform of the measured waveforms. (c) The frequency spectrum of $U_{i,1}(f)$. (d) The frequency spectrum of $U_{r,1}(f)$. (e) The frequency spectrum of $U_{r,2}(f)$. (f) The frequency spectrum of $U_{t,2}(f)$. (g) The theoretical and measured frequencies of spatiotemporal scattered waves in subluminal, superluminal and interluminal regimes. The theoretical calculations are presented as continuous curves, and the experimental results are presented as discrete points [normalized reflected frequency $f_{r,1}/f_{i,1}$ ($f_{r,2}/f_{i,1}$) is marked in red (blue), and the normalized transmitted frequency $f_{r,1}/f_{i,1}$ is marked in green]. (h) The theoretical and measured scattering coefficient corresponding to the scattered waves in (g) [the reflection coefficient $R_1$($R_2$) is marked in red (blue), and the transmission coefficients $T$ is marked in green]. Insets in (h) show the FDTD simulated result of the spatiotemporal scatterings in subluminal (leftmost inset), superluminal (rightmost inset) and interluminal (middle inset) regimes.

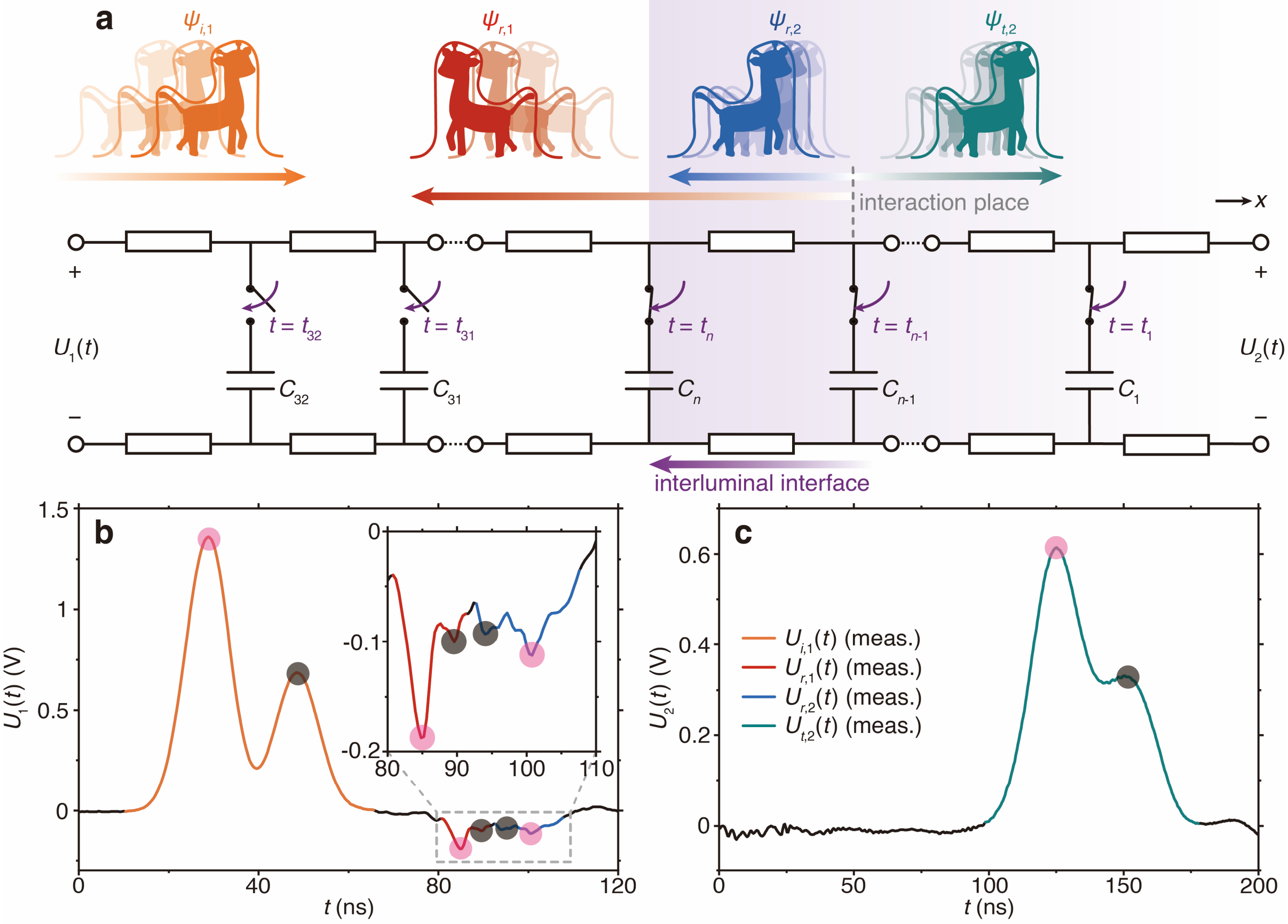

**Fig. 3 | Analysis of the causality of the scattered waves at interluminal interface.** (a) Schematic of the causality of the wave scattering in interluminal regime. An asymmetric signal $\psi_{i,1}$ is input from port 1 [the port voltage is $U_1(t)$], and interluminal interface is moving toward the negative *x*-axis. The interaction between $\psi_{i,1}$ and the interluminal interface results three scattered waves $\psi_{r,1}$, $\psi_{r,2}$, and $\psi_{t,2}$. The spatial sequence of the $\psi_{r,1}$ (marked in red) is opposite to the $\psi_{i,1}$ (marked in orange), while that of the $\psi_{r,2}$ (marked in blue) is the same as the $\psi_{i,1}$. (b, c) The experimental results of the causality of the spatiotemporal scattering. (b) The temporal variation of voltage wave $U_1(t)$ at port 1. The incident waves $U_{i,1}(t)$ (highlighted in orange) are set to a pair of asymmetric Gaussian pulses with different magnitudes, where the first pulse has a higher magnitude (marked by pink dot), while the second one has a lower magnitude (marked by black dot). For the first received reflection voltage wave $U_{r,1}(t)$ (highlighted in red), the pulse with the higher magnitude arrived first, followed by the lower one (i.e., the pink dot appears before the black dot). For the second received reflection voltage wave $U_{r,2}(t)$ (highlighted in blue), the pulse with the lower magnitude arrived first, followed by the higher one (i.e., the pink dot appears after the black dot). (c) The temporal variation of voltage wave $U_2(t)$ at port 2. For the received transmission voltage wave $U_{t,2}(t)$ (highlighted in green), the pulse with the higher magnitude arrived first, followed by the lower one (i.e., the pink dot appears before the black dot).